# Optical frequency comb double-resonance spectroscopy of the 9030-9175 cm$^{-1}$ states of ethylene

Adrian Hjältén[1], Vinicius Silva de Oliveira[1], Yuan Cao[1], Isak Silander[1], Kevin K. Lehmann[2], and Aleksandra Foltynowicz[1,*]

[1] Department of Physics, Umeå University, 901 87 Umeå, Sweden

[2] Departments of Chemistry & Physics, University of Virginia, Charlottesville, VA 22904, USA

*aleksandra.foltynowicz@umu.se

**Abstract**

We use optical-optical double-resonance (OODR) spectroscopy to measure for the first time hot-band transitions of ethylene ($C_2H_4$) between states in the 3000 cm$^{-1}$ and 9000 cm$^{-1}$ energy ranges. A 3.2 μm continuous wave (CW) pump is used to populate selected states in the $\nu_9$ vibrational mode. The sub-Doppler OODR transitions are then probed with two different cavity-enhanced probes tunable around 1.7 μm: a frequency comb probe that allows for broadband measurements and simultaneous detection of many OODR lines, and a CW probe that measures individual lines with higher signal-to-noise ratio and better frequency accuracy. We report center frequencies and relative intensities of 90 ladder-type hot-band transitions from three different states in the $\nu_9$ vibrational mode. We exploit combination differences and measurements of polarization-dependent intensity ratios to determine the final state rotational quantum numbers $J$. Comparison to theoretical predictions from ExoMol allows tentative assignments for 28 transitions. We report improved center frequencies for the three pump transitions in the $\nu_9$ band. Furthermore, we observe 18 sub-Doppler V-type transitions from the depleted ground state to the 6000 cm$^{-1}$ region and assign 14 of them to the variational line list of Mraidi S *et al.* [J Quant Spectros Radiat Transfer. 2023;310:108734 doi:10.1016/j.jqsrt.2023.108734].

**1. Introduction**

Ethylene, $C_2H_4$, is a component of the Earth's atmosphere, where it is produced by both natural and anthropogenic sources and impacts the concentration of other components, e.g. ozone [1, 2]. In addition, it has been detected on several planetary bodies in the solar system [3-5]. Spectroscopic detection of ethylene is therefore important for remote sensing, both on Earth and other celestial bodies.

The molecule possesses $D_{2h}$ point symmetry and 12 vibrational modes (5 IR active, 6 others Raman active), but has a less clear polyad structure than, e.g., $CH_4$ [6, 7]. This together with numerous resonance couplings, gives rise to highly congested and complex spectra that are challenging to model. Ethylene has been quite well studied in the 600-1500 cm$^{-1}$ region of strong fundamental bands [8-11], also including some hot bands [12, 13] and bands of the $^{13}$C-isotopologues [13, 14]. The CH-stretch region around 3000 cm$^{-1}$ was studied using tunable laser spectroscopy by Pine [15]; this is the source of the HITRAN data in this region. More recent works have reported Hamiltonian parameters of bands in this region [16-19], as well as empirical and calculated line lists [20-22]. Optical-optical double resonance (OODR) spectroscopy of $C_2H_4$ has been performed using mid-infrared (MIR) CW lasers in a molecular jet [23-25] to address hot band transitions reaching vibrational states around 4000 cm$^{-1}$. Several high-resolution studies of the overtone bands in the near-infrared have been conducted using narrow-band lasers around 6000-6200 cm$^{-1}$ [26, 27], and Fourier transform infrared spectroscopy (FTIR) up to 7900 cm$^{-1}$ [20, 28-30]. Some of these works employed molecular beams [28], or slit-jet cooling [29, 31] to decongest the spectrum.

A global line list in the range 0-6700 cm$^{-1}$ based on variational calculations was presented by Rey *et al.* [7] and included in the TheoReTs database [32]. This line list was extended up to 9000 cm$^{-1}$ by Mraidi *et al.* [33] and compared to FTIR spectra, where the agreement in line positions with the experiment was found to be ~1 cm$^{-1}$. Ben Fathallah *et al.* [30] presented an empirical line list in the 5800-6400 cm$^{-1}$ range based on an FTIR spectrum, which was partly assigned to the calculated line list of Mraidi *et al.* [33]; only about 15% of the observed lines could be assigned due to the strong congestion of the spectrum.

The HITRAN2020 database [34] contains C$_2$H$_4$ line lists only in two MIR regions between 620-1520 cm$^{-1}$ and 2920-3240 cm$^{-1}$. No updates to the ethylene line list were made in the 2024 update of HITRAN [35]. On the other hand, the ExoMol database [36] provides a line list that reaches up to 7000 cm$^{-1}$ in wavenumber and contains rotational-vibrational levels up to 16000 cm$^{-1}$, obtained from *ab initio* dipole moment and potential energy surfaces, where the latter was refined based on empirical data. The ongoing MARVEL project which replaces theoretical term values with ones derived from experimental spectra will improve the energy levels up to 6300 cm$^{-1}$ [37]. Since the Mraidi *et al.* line list [33] does not contain hot bands starting from levels above 1950 cm$^{-1}$, the only available predictions for hot bands starting from levels in the 3000 cm$^{-1}$ range is the purely calculated line list from the ExoMol database.

Here we use optical-optical double-resonance spectroscopy with a 3.2 μm CW pump and two cavity-enhanced probes centered at around 1.7 μm: a frequency comb and a CW external cavity diode laser (ECDL) [38] to measure ground-state and hot-band transitions of ethylene in the 5880-6080 cm$^{-1}$ range. The comb probe allows simultaneous detection of a large number of OODR lines, while the CW probe enables the measurement of individual lines with higher signal-to-noise ratio and improved frequency accuracy. Figure 1a) depicts schematically the vibrational bands addressed by the CW pump (red), comb probe (green) and CW probe (blue). Without the pump excitation, the probe induces Doppler-broadened transitions from the ground state to the overtone region at 6000 cm$^{-1}$. The Doppler-broadened spectrum consists of a congested abundance of overtone and combination bands, the strongest being $\nu_5+\nu_9$, $\nu_5+\nu_{11}$, $\nu_2+\nu_3+\nu_{11}$, $\nu_1+\nu_{11}$ and $\nu_1+\nu_2+\nu_{12}$ [6]. When the pump is tuned to a selected transition in the $\nu_9$ band, the probe excites additional sub-Doppler ladder-type transitions from the pumped $\nu_9$ levels around 3000 cm$^{-1}$ to the 9000 cm$^{-1}$ region. Doppler-broadened lines sharing the lower state with the pump transition will also display sub-Doppler V-type features at their centers caused by lower state population depletion by the pump.

Figure 1b) shows the details of the rotational structure of the states probed in this work. Using the comb probe, we observe ladder-type hot-band transitions starting from the upper states of the P(4,A$_g$), Q(4,A$_g$) and R(4,A$_g$) transitions of the $\nu_9$ fundamental band with rotational quantum numbers $J'$ = 3 – 5, reaching rovibrational states in the 9000 cm$^{-1}$ region and V-type transitions from the vibrational ground state with $J''$ = 4. V-type transitions thus reach final states with $J$ = 3 – 5, while ladder-type transitions reach levels with a final $J$ = 2 – 6, where $J$ = 3 – 5 can be reached by more than one pump-probe combination. Such so-called combination differences allow for restricting the possible final state $J$-numbers of ladder-type transitions. In particular, final states probed for all three pump transitions, or at least for the P(4,A$_g$) and R(4,A$_g$) pump transitions, must have $J$ = 4, where the lacking observation in the Q(4,A$_g$)-pumped spectrum might be due to the transition falling below the experimental detection limit. In addition, we compare line intensities measured at parallel and perpendicular pump-probe polarization, which allows for transition branch assignment [39]. For the R(4,A$_g$) pump transition, we improve the measurement sensitivity by using the CW probe. We confirm the existence of a ladder-type transition predicted based on a combination difference, but too weak to be detected by the comb probe. We determine the intensity ratios of selected ladder-type transitions with better precision, allowing for more conclusive branch assignment than possible in the comb spectra. Using the CW probe, we also determined the center frequency of the $\nu_9$ R(4,A$_g$) transition with higher accuracy than available from the literature.

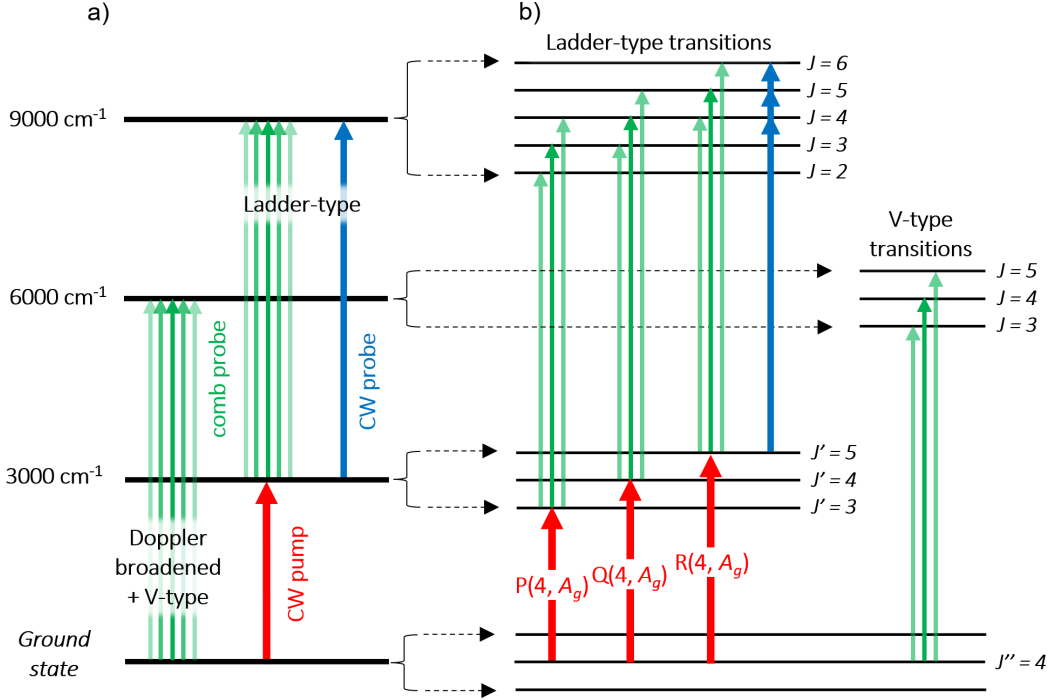

Figure 1. A schematic depiction of the transitions addressed by the CW pump (red), comb probe (green) and CW probe (blue). a) Vibrational band regions addressed by the three light sources. b) Rotational levels addressed by the pump and probes in a ladder-type configuration for the three pump transitions in this work, showing combination differences for final states with $J = 3 – 5$ in the 9000 cm$^{-1}$ range. V-type features always appear on transitions from the state depleted by the pump and reach rotational levels with $J = 3 – 5$ in the 6000 cm$^{-1}$ region.

## 2. Spectral acquisition

The experimental setup for combined frequency comb and CW OODR spectroscopy is described in detail in Ref. [38] and is briefly summarized below. The 3.2 μm CW pump is the idler of a singly-resonant optical parametric oscillator (TOPTICA, TOPO) phase-locked to a home-built 125 MHz MIR frequency comb [40]. The comb probe is a 250 MHz Er:fiber optical frequency comb (Menlo Systems, FC-1500-250-WG) with the output spectrum shifted by a polarization-maintaining microstructured silica fiber to operate around 1.7 μm with 180 cm$^{-1}$ of a simultaneous bandwidth at the -10 dB level. The CW probe is an external cavity diode laser (Sacher, LION model P-1650) with output tunable between 1.6–1.75 μm.

The probe beam is mode-matched to the TEM$_{00}$ mode of a 60-cm-long Fabry-Perot cavity with a finesse around 1000. The cavity mirrors have 1-m radius of curvature, resulting in a probe beam radius at the waist of 0.5 mm. The comb or CW probe are selected by switching the fiber leading to the free space mode-matching telescope. The pump beam is combined collinearly with the probe beam using a dichroic mirror and focused on the center of the cavity with a radius of 0.7 mm so its Reyleigh range of 45.8 cm is equal to the probe's. The pump beam is not resonant with the cavity.

The frequency comb probe is locked to the cavity using the Pound-Drever-Hall (PDH) method [41], with feedback to the repetition rate, $f_{rep}$, using an error signal from a part of the spectrum selected using a grating, referred to as the PDH locking point. The carrier-envelope-offset, $f_{ceo}$, is RF-locked to a frequency that maximizes the spectral bandwidth of the cavity transmission. A third locking loop stabilizes the sample cavity length via a piezo-electric transducer (PZT) and thereby allows absolute stabilization and tuning of $f_{rep}$.

The comb probe beam transmitted through the cavity is coupled to a Fourier transform spectrometer (FTS) with a nominal resolution matched to $f_{rep}$ and auto-balanced detection. A frequency-stabilized

HeNe reference laser (Sios, SL/02/1) with a wavelength $\lambda_{ref} \approx 632.991$ nm is used to calibrate the optical path difference (OPD). We use the sub-nominal sampling-interleaving method [42, 43], which requires matching the spectral sampling points to the comb mode frequencies. We achieve this by iteratively adjusting the reference laser wavelength $\lambda_{ref}$ assumed in post processing to minimize the instrumental line shape (ILS) of 4 strong ladder-type transitions in each measurement. We take the optimum $\lambda_{ref}$ to be the mean of the 4 optima, and the uncertainty to be the standard deviation of their scatter.

In the measurement series using the comb probe, we acquired spectra at 132 different $f_{rep}$ values spaced by 2.75 Hz, corresponding to 2 MHz shifts of the comb modes in the optical domain. For each $f_{rep}$, we recorded a pair of spectra with and without pump excitation by opening and closing a shutter positioned in the pump beam. The $f_{rep}$ was stepped and after completing a full $f_{rep}$ scan in about 15 minutes, the pump polarization was rotated by 90° by turning a half-wave plate. The acquisition series was repeated, stepping $f_{rep}$ in the opposite direction. For averaging, we made several back-and-forth scans of $f_{rep}$.

When using the CW probe, the Er:fiber frequency comb is used as frequency reference and its $f_{ceo}$ and $f_{rep}$ are RF stabilized. The CW probe is PDH-locked to the enhancement cavity via feedback to the laser current and the beat note between the CW probe and the Er:fiber frequency comb is stabilized at 61 MHz by controlling the enhancement cavity length. The CW probe is scanned across transitions in steps of 180 kHz by stepping the Er:fiber frequency comb $f_{rep}$ in 0.25 Hz steps, which in turn changed the frequency of the CW probe via the feedback loops. The acquisition time for each sampling point is 0.6 s. The CW probe is scanned in both directions, i.e. while increasing and decreasing the optical frequency. A complete scan over the ± 23 MHz range lasts roughly 2.5 min. More details about the frequency stabilization are given in Ref. [38]. The CW probe beam transmitted through the sample is detected directly after the cavity using an InGaAs detector (Thorlabs, PDA20CS-EC). The pump is amplitude modulated at around 500 Hz using a mechanical chopper and the modulated cavity transmission probe signal is detected using a lock-in amplifier (Stanford Research, SR830).

We made three measurement series with the comb probe, pumping the $^RP_{0,4}(4,A_g)$, $^RQ_{0,4}(4,A_g)$ and $^RR_{0,4}(4,A_g)$ transitions of the $\nu_9$ band. We selected this set of pump transitions since they are three strong lines that share the lower state and do not overlap with surrounding transitions. We also wanted to avoid transitions to levels with too high $J$-numbers as that is detrimental to the contrast in intensity ratios for different relative pump-probe polarizations [39]. The pump was sequentially locked to the center wavenumbers of the pump transitions taken from HITRAN2020 [34] via locking to the MIR frequency comb. Of the other available line lists in the 3000 cm$^{-1}$ range, Ref. [21] claims worse line center accuracy than HITRAN and the discrepancies of ExoMol and Ref. [20] compared to HITRAN are larger than the uncertainties of the latter. Unlike in our previous work on methane [38, 44], we could not observe Lambs dip in the pump transitions since they were about one order of magnitude weaker than the transitions we pumped before.

The experimental details of the comb probe measurements are summarized in Table 1. The assignments of the pump transitions are given in the form ($J,C_{tot},N$), where $J$ is the total angular momentum quantum number, $C_{tot}$ the ro-vibrational symmetry, and $N$ the ranking number. The sample pressure in all measurements was 100 mTorr, which we chose as a trade-off between maximizing the signals of OODR ladder-type transitions while still limiting the absorption of Doppler-broadened ground-state transitions to be able to see the V-type features at their centers.

Table 1. The conditions of the three measurements with the frequency comb probe. Columns: 1. Measurement number, 2. Pump transition assignment and wavenumber from HITRAN2020 [21], 3. Probe spectral coverage at the -10 dB level, 4. PDH locking point of comb $f_{rep}$, 5. Pump power incident on the sample, 6. Pump transmission on resonance with the pump line, 7. Number of $f_{rep}$ scans for parallel (∥) and perpendicular (⊥) pump-probe polarization.

| Measurement | Pump transition and wavenumber [cm$^{-1}$] | Coverage [cm$^{-1}$] | PDH locking point [cm$^{-1}$] | Pump power at sample [mW] | On-res. transmission [%] | # scans ∥ / ⊥ polarization |
|---|---|---|---|---|---|---|
| #1 | P(4,$A_g$) (3,$A_u$,8)←(4,$A_g$,1) 3101.043322 | 5910-6080 | 6024 | 685 | 94 | 10 / 9 |
| #2 | Q(4,$A_g$) (4,$A_u$,6)←(4,$A_g$,1) 3109.715028 | 5920-6090 | 6015 | 693 | 90 | 6 / 5 |
| #3 | R(4,$A_g$) (5,$A_u$,10)←(4,$A_g$,1) 3116.658932 | 5900-6080 | 6006 | 840 | 90 | 9 / 9 |

We used the CW probe to measure five selected R(4,$A_g$)-pumped ladder-type transitions with parallel and perpendicular relative pump-probe polarization, for determination of the polarization dependent intensity ratios with better precision than with the comb probe. These measurements also allowed for a more accurate determination of the wavenumber of the $\nu_9$ R(4,$A_g$) pump transition than listed in HITRAN2020, after we observed splitting of the OODR transitions (see Section 3.1.1), which is an indication of a detuning from the pump transition. Compared to the comb probe measurement conditions, we used a lower pump power of around 15 mW to reduce the power broadening, and the sample pressure was in the 10 mTorr range.

## 3. Analysis

### 3.1. Comb probe measurements

To detect the OODR transitions and to fit the ladder-type transitions we normalized the spectra acquired with pump on to the background spectra acquired with pump blocked and interleaved them to a point spacing of 2 MHz, as described in Refs. [38, 44]. The normalization step largely cancels the Doppler-broadened ground state transitions and the shape of the comb envelope. We detected the remaining sub-Doppler OODR V-type and ladder-type features using a peak finding routine described in Ref. [44].

To retrieve the enhancement factor of the cavity and to fit the V-type transitions we interleaved sets of spectra without background normalization following the same procedure as in Refs. [38, 44] using the empirical line list of Ben Fathallah *et al.* [30] (instead of HITRAN) to model the Doppler-broadened $C_2H_4$ ground state absorption and dispersion when removing the baseline. We retrieved the wavenumber-dependent cavity-enhancement factor by fitting a model of the cavity transmission function [45] to the Doppler-broadened lines in the baseline-corrected spectrum interleaved to 20 MHz sample point spacing. The procedure is identical to that in Ref. [38] except the Doppler-broadened cavity enhanced spectrum was simulated using the Ben Fathallah experimental line list [30]. The cavity enhancement factors retrieved from the parallel and perpendicular pump polarizations were found to differ insignificantly for a given measurement set and for simplicity we used only the results from the parallel case. For fitting of the V-type transitions, we interleaved the baseline-corrected spectra to a point spacing of 2 MHz.

### 3.1.1. Fitting of ladder-type transitions

Figure 2 shows three ladder-type lines (black) observed in the three comb probe measurements, all measured with parallel relative pump-probe polarization. The line fits are shown in red and the bottom panels display the fit residuals.

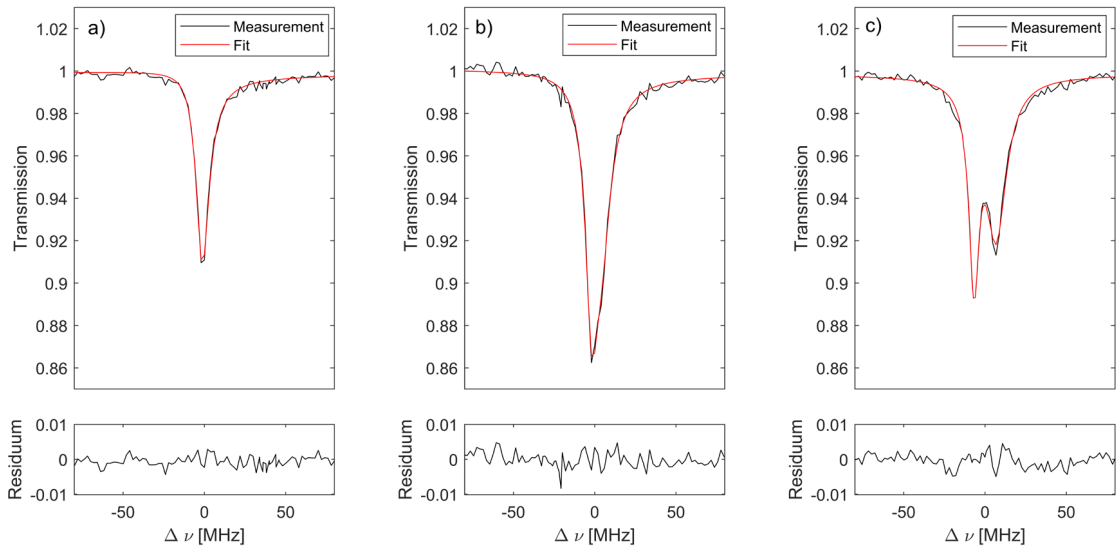

Figure 2. Ladder-type transitions (black) observed with the comb probe and parallel relative pump-probe polarization at a) 6041.056365(2) cm$^{-1}$ in measurement #1, at b) 6026.472250(3) cm$^{-1}$ in measurement #2 and at c) 6025.440881(3) cm$^{-1}$ in measurement #3, as well as the corresponding line fits (red). The bottom panels show the fit residuals. The split peak is clearly visible in panel c) while the line in panel b) displays a slight asymmetry; both effects are accounted for by the model.

In measurement #1, shown in Figure 2a), we fitted the ladder-type lines as described in Refs. [38, 44] where the lines were modeled as a sum of a Lorentzian component for the sub-Doppler part and a broader Gaussian component accounting for collision-induced velocity redistribution. The free parameters were a common center frequency of both components, the integrated absorption and width of the Lorentzian component, and the comb-cavity phase offset. We determined the ratio of the integrated absorption of the Gaussian to the Lorentzian component and the Gaussian width for a single line with a good SNR and fixed it to these values for all remaining lines. Depending on how pronounced the baseline structure was around a given line, either a 1$^{st}$ or 3$^{rd}$ order polynomial baseline was included in the fit, and the fit window width was chosen between ±100 MHz and ±450 MHz around the line center.

In measurement #3, all OODR lines were found to be split into two peaks separated by approximately 14 MHz, as seen in Figure 2c). This suggests that the pump laser frequency was detuned from resonance of the pump transition, which leads to frequency shifts in OODR absorption features in opposite directions for the probe co- and counterpropagating to the pump inside the cavity. We did not find any indication that the pump laser frequency had been mistakenly detuned from the R(4,A$_g$) transition wavenumber provided in HITRAN and therefore suspected an inaccuracy in the HITRAN value. We modified the fitting routine to accommodate the split OODR features by incorporating a pair of Lorentzian-Gaussian line shapes with identical integrated absorption, separated by offsets equal in magnitude but opposite in sign from the center frequency of the transitions. In addition to the center frequency, the common intensity of the peaks and the frequency split between them, we fitted the two Lorentzian widths separately since they should differ depending on the propagation direction relative to the pump [46]. We found the mean frequency splitting for 10 lines with the best SNR to be 13.9(3) MHz and the mean Lorentzian widths to be 5.81(14) MHz and 4.21(15) MHz, yielding a ratio of 1.38(6), with the higher frequency component being broader. The uncertainties are the standard deviations of values. The theoretical ratio of the widths is 1.67 (5:3) with the copropagating component being broader [46]. The pump detuning from the transition center is obtained by dividing the frequency split by a factor of 2 to obtain the shift from the center frequency and then by the ratio of the probe to the pump frequencies, which is in the range 1.89 – 1.96. The frequency detuning of the co- and counterpropagating peaks determines the sign of the pump detuning with respect to the center of the transition. Our measurements indicate that the center of the R(4,$A_g$) transition is 3.56(7) MHz lower than the HITRAN value, which is comparable to the 3 MHz uncertainty claimed in HITRAN for this transition.

In measurement #2, while the lines were not split, the retrieved Lorentzian widths were 60% larger than in measurement #1 and for a few ladder-type lines with high SNR, the line shape appeared asymmetric, as in Figure 2b). This indicated that the co- and counterpropagating line shapes were slightly split also here. We used the same fitting routine as for measurement #3, and we obtained a mean frequency split of 5.75(50) MHz and a mean ratio of the Lorentzian widths of 1.53(16) from fits to six lines with sufficient SNR for the asymmetry to be apparent. The splitting indicates that frequency of the Q(4,$A_g$) transition is 1.49(13) MHz lower that stated in HITRAN.

### 3.1.2. Fitting of V-type transitions

In measurement #1 we fitted the V-type transitions as described in Ref. [38]. We first canceled the neighboring Doppler-broadened lines by division with a cavity transmission model based on the Ben Fathallah *et al.* empirical line list [30] and fitted out the Doppler-broadened line displaying the V-type using a pure Gaussian model with a full width at half maximum of 420 MHz in the cavity transmission function. We fitted the V-type feature remaining in the spectrum in a window of ±200 MHz using the same model as for ladder-type transitions but inverting the sign of both the absorption and dispersion components in the transmission function. The fit included a 3$^{rd}$ order polynomial baseline for all lines. In measurements #2 and #3, we again modified the model to include two V-type features with the same intensity but different width. We fixed the values of the frequency split and the ratio of the Lorentzian widths to the mean values obtained from the ladder-type fits in the corresponding measurement, described in Section 3.1.1. Figure 3 displays the R(4,$A_g$) V-type transition in the $\nu_5+\nu_{11}$ band observed in the three measurements (black) at parallel relative pump-probe polarization together with the line fits (red). The lower panels show residuals.

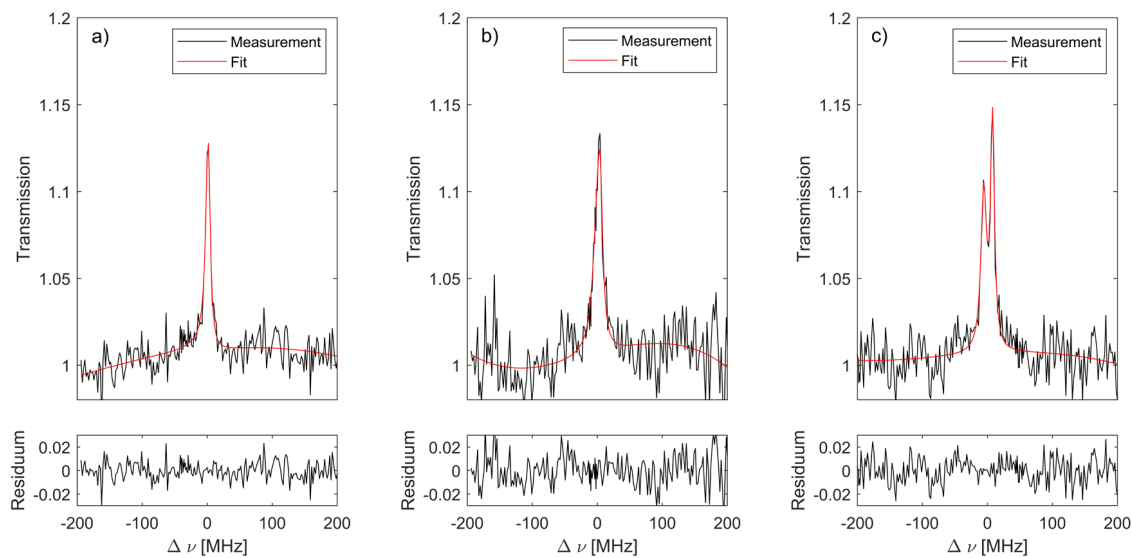

Figure 3. A R(4,$A_g$) V-type transitions in the $\nu_5+\nu_{11}$ band (black) observed at 6006.522086(6) cm$^{-1}$ using the comb probe with parallel relative pump-probe polarization in a) measurement #1 and b) measurement #2 and c) measurement #3, as well as the corresponding fits (red). The bottom panels show residuals.

## 3.2. CW probe measurements

To more accurately determine the center frequency of the $\nu_9\,R(4,A_g)$ pump transition, we intentionally detuned the pump frequency by ~5.5 MHz from the transition center determined in the comb probe measurement to increase the separation between the co-/counter propagating peaks. We scanned the CW probe over the strongest ladder-type transition at 6044.25967246(9) cm$^{-1}$ 41 times and retrieved the split by fitting a model containing two Lorentzian line shapes and a 2$^{nd}$ order polynomial baseline, shown in Figure 4a). The fit parameters were the center frequency, width and intensity of the Lorentzian line shapes, in addition to the baseline coefficients. The width of the narrower peak is 1.07(4) MHz, and the width of the broader peak is 1.51(5) MHz. This is again comparable to the theoretical width ratio of 3:5. Single scans were fit individually to compute statistics instead of averaging the measurements.

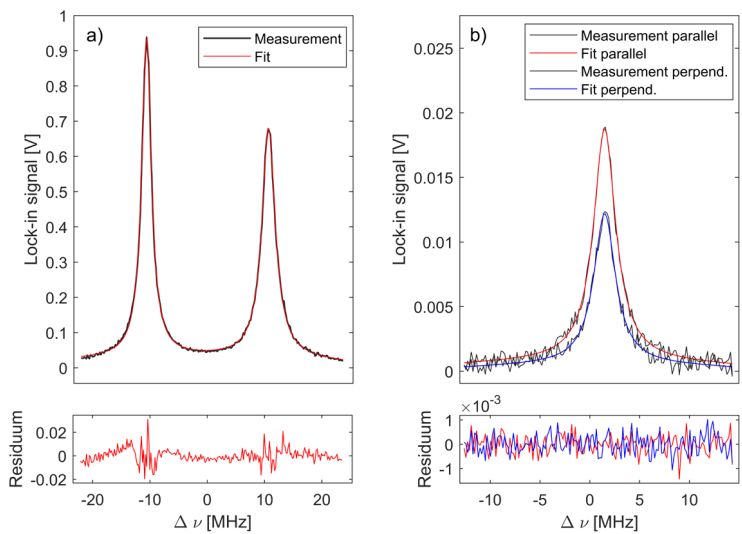

Figure 4. Ladder-type transitions measured using the CW probe. a) Measurement (black) and fit (red) of the R(5,A$_u$) ladder-type transition at 6044.25967246(9) cm$^{-1}$ with parallel relative pump-probe polarization, used for retrieving the center frequency of the $\nu_9\,R(4, A_g)$ pump transition. The pressure was 15 mTorr and the pump was detuned from the center of transition by ~5.5 MHz. b) Measurement of the P(5,A$_u$) line at 5901.2634675(1) cm$^{-1}$ at 14 mTorr (black) and Lorentzian fits for parallel (red) and perpendicular (blue) relative pump-probe polarization. The line was not observed in the comb-OODR measurements but was predicted via combination differences.

From the retrieved mean frequency splitting between the peaks of 21.262(15) MHz, and the CW pump laser frequency calculated using the beat between the CW pump and the MIR frequency comb, we obtained a frequency of the $\nu_9\,R(4,A_g)$ transition 3.265(4) MHz lower than the HITRAN value, where the uncertainty is the standard deviation of separation between the two peak frequencies determined in the different scans. This value is 295 kHz lower than the value retrieved from the comb probe ladder-type fits (in Section 3.1.1). However, we deem the result from the CW measurement more accurate since the two peaks are narrower and better resolved than in the comb measurement shown in Figure 2c) due to lower pump power and consequently reduced power broadening in the CW measurement.

Using the CW probe, we also measured five ladder-type transitions at parallel and perpendicular relative pump-probe polarization to more accurately determine the polarization dependent intensity ratios. Figure 4b) shows single scans at two relative pump-probe polarizations that illustrate the change in intensity of the line at 5901.2634675(1) cm$^{-1}$ (black). The pump frequency is here adjusted to the center of the $\nu_9\,R(4,A_g)$ transition. The line was fitted using the same routine as above but with a single Lorentzian, and the fits are shown for the parallel (red) and perpendicular (blue) polarization configuration. The mean width of the line is 1.37(7) MHz and the intensity ratio is 1.52(5), which indicates it is a P(5,A$_u$) transition. Section 4.3 provides a detailed discussion and comparison of the polarization dependent intensity results obtained from CW probe and comb probe measurements.

## 4. Results

### 4.1. Line parameters

We calculated the line centers of the transitions observed in the comb-OODR measurements as the mean of the values obtained from fits to the spectra at the two relative pump-probe polarizations, weighted by the inverse square of the uncertainties, obtaining the total uncertainty by error propagation. We calculated the total values of the integrated absorption as ($\alpha_\parallel$+2$\alpha_\perp$)/3, where $\alpha_\parallel$ and $\alpha_\perp$ are the integrated absorption of the Lorentzian fit component with parallel and perpendicular relative pump-probe polarization respectively, and propagated the uncertainties accordingly. We also calculated the intensity ratios $\alpha_\parallel$/$\alpha_\perp$, propagating the uncertainties from both components. The line positions, integrated absorptions, and the intensity ratios for all detected ladder-type lines are listed in Tables A 1 – A 3 in the Appendix.

We estimated the total uncertainty in the retrieved line positions as originating from two sources, the line fits and the optimization of the FTS CW reference laser wavelength $\lambda_{ref}$. The fit uncertainties varied between 60 kHz and 4.5 MHz depending on the SNR of the line. We estimated the contribution from the $\lambda_{ref}$ optimization by fitting 4 ladder-type transitions with good SNR in each measurement in spectra analyzed with a range of $\lambda_{ref}$ values around the optimum and multiplying the largest observed $\lambda_{ref}$-dependence of the line center by the uncertainty in $\lambda_{ref}$, estimated as described in Section 2. This contribution was 30-70 kHz depending on the measurement.

When using the CW probe, we calculated the mean of the center frequencies fitted to all acquired scans and the mean line areas from the scans at the two polarization settings. We took the frequency uncertainties to be the standard error of the mean. The polarization ratio determining the branch of the transitions is calculated as the ratio of the mean areas at parallel and perpendicular polarization and the uncertainty is the standard deviations of the line areas combined in quadrature.

We did not find any reported values of the pressure shift in the wavenumber region studied in this work and we did not correct the reported center frequencies for the pressure shift, nor do we have an estimate of the uncertainty stemming from it.

### 4.2. Combination differences and improved pump transition frequencies

We calculated the final state term values for each observed ladder-type line by summing their transition wavenumbers, the wavenumbers of the pump transitions from HITRAN, where we applied the correction from Section 3.2 to the $\nu_9$ R(4,A$_g$) transition, and the MARVELized ground state term value of 18.20573(7) cm$^{-1}$ [37]. We identified 16 combination differences by grouping final state term values that differed by less than 5 × 10$^{-4}$ cm$^{-1}$ (15 MHz). Some systematic disagreement in the term values was observed, which is expected due to the already observed offset in pump transition frequencies in HITRAN. We hence used the combination differences to obtain corrections to the wavenumbers of the $\nu_9$ P(4,A$_g$) and Q(4,A$_g$) pump transitions. We selected 6 combination differences involving 13 ladder-type transitions with good SNR and sub-MHz center frequency uncertainty. The frequency of the $\nu_9$ R(4,A$_g$) line was fixed to the value determined in the CW probe measurement while the other two were varied using the MATLAB™ function "fminsearch" to minimize the standard deviation of the scatter of the term values of each combination difference around their mean. The standard deviation of the remaining scatter was taken as the uncertainty in the correction to the pump transitions. In this way, we determined a correction of -2.0(3) MHz for the Q(4,A$_g$) transition, while the correction to the P(4,A$_g$) transition was smaller than the uncertainty and hence omitted. Again we note that the correction to the Q(4,A$_g$) transition obtained here differs from that retrieved from spitting of the ladder-type lines in the comb-OODR spectrum, but we consider the latter value less reliable due to the splitting in the ladder-type being barely detectable (recall Figure 2b)). The pump transition frequencies are summarized in Table 2 and compared to the values from HITRAN and the experimental line list of Ben Fathallah *et al.* [21]. We used these corrected pump transition frequencies when calculating the upper state term values reported in this work.

Table 2. The pump transition wavenumbers retrieved in this work compared to HITRAN and Ben Fathallah *et al.* [21]. The last column indicates whether the corrections to the HITRAN values were obtained by measurement with the CW probe (CW) or by optimizing the agreement in combination differences (CD).

| $\nu_9$ band transition | Line center from this work [cm$^{-1}$] | Line center HITRAN [34] [cm$^{-1}$] | Line center Ref. [21] [cm$^{-1}$] | Verification method |
|---|---|---|---|---|
| P(4,A$_g$) | 3101.04332(1) | 3101.0433(1) | 3101.0435(2) | CD |
| Q(4,A$_g$) | 3109.71496(1) | 3109.7150(1) | 3109.7150(2) | CD |
| R(4,A$_g$) | 3116.65882309(13) | 3116.6589(1) | 3116.6591(2) | CW |

### 4.3. Polarization-dependent intensity ratios

The theoretical intensity ratios of ladder-type lines measured with the pump polarization parallel and perpendicular to the probe polarization are described in Ref. [39] and reproduced in Table 3 for the three pump transitions in this work.

Table 3. The theoretical intensity ratios of ladder-type transitions observed with parallel and perpendicular relative pump-probe polarization when pumping P(4), Q(4), and R(4) transitions, taken from Ref. [39].

|  | Probing P transition | Probing Q transition | Probing R transition |
|---|---|---|---|
| Pumping P(4) transition | 1.1690 | 0.8125 | 1.0682 |
| Pumping Q(4) transition | 0.5455 | 2 | 0.7500 |
| Pumping R(4) transition | 1.5089 | 0.4640 | 1.2739 |

The accuracy of the intensity-based branch assignment for ladder-type transitions measured using the comb probe was limited by the SNR. Figure 5 shows by the open black markers the ratios of intensities of the ladder-type lines measured using the comb probe with parallel and perpendicular pump-probe polarization in comb probe measurement #3, i.e., pumping the $\nu_9$ R(4,A$_g$) transition. For clarity, the data are plotted against line number in order of increasing transition wavenumber (same as in Table A3) ensuring that the points are evenly spaced along the x-axis. The error bars are the 1σ uncertainties. The horizontal lines show the predicted intensity ratios of probe transitions in the P- (blue dotted), Q- (green dash-dotted) and R-branches (red dashed). The precision of the measurement does not allow for unambiguously distinguishing between the P and R transitions that have similar intensity ratios. For comparison, the solid red markers show the intensity ratios for the five transitions measured using the CW probe, which in general have lower uncertainties because of the higher SNR. The transition wavenumbers and polarization dependent intensity ratios of these five ladder-type transitions are shown in Table 4. All lines except the line at 6044.25967246(9) cm$^{-1}$ (measured as described in Section 3.2) were measured with the pump tuned to the center of the R(4,A$_g$) pump transition. By comparison with the predicted intensity ratios, the branch of the transitions can unambiguously be concluded and thus the upper state *J*-number is given in the last column. Note that the first line at 5901.2634675(1) cm$^{-1}$ was too weak to be observed in the comb measurement but its presence was predicted based on a combination difference involving lines in measurements #1 and #2. Its detection using the CW probe thus confirms the upper state *J*-number of those two lines as well.

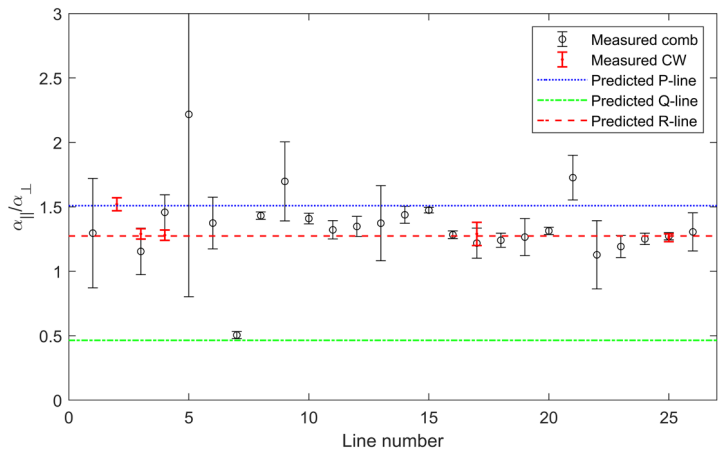

Figure 5. The ratios of integrated absorption retrieved with parallel and perpendicular pump-probe polarization in comb measurement #3 (open black markers) and using the CW probe (solid red markers). The predicted ratios for probe lines in the P- (blue dotted), Q- (green dash-dotted) and R-branches (red dashed) are shown as horizontal lines. The x-axis shows the number of the line in order of increasing center wavenumber.

Table 4. The transition wavenumbers, the intensity ratios of parallel-to-perpendicular pump-probe polarization, and the assigned upper state $J$-number for the five ladder-type transitions probed with the CW laser and pump on the $\nu_9$ R(4,A$_g$) transition. The frequency uncertainty is reported as the standard error of the mean.

| Transition wavenumber [cm$^{-1}$] | Intensity ratio $\alpha_\parallel/\alpha_\perp$ | Upper state $J$-number |
|---|---|---|
| 5901.2634675(1) | 1.52(5) | 4 |
| 5912.9344311(1) | 1.29(4) | 6 |
| 5920.8322341(3) | 1.28(4) | 6 |
| 6039.7881914(2) | 1.29(9) | 6 |
| 6044.25967246(9) | 1.26(3) | 6 |

## 4.4. Ladder-type assignment

We fitted a total of 89 ladder-type transitions in the comb-OODR spectra, 30 in measurement #1, 34 in measurement #2, and 25 in measurement #3. The 16 combination differences involved 37 of the observed ladder-type transitions, of which 28 could be concluded to have final state with $J = 4$, when considering also the line measured only with the CW probe. While we cross-checked assignments with the polarization-dependent intensity ratios, the absorption sensitivity in the comb-OODR data were deemed insufficient to provide conclusive branch assignments (recall Figure 5). Out of the lines whose $J$-numbers were confirmed by a combination difference, only 70% had the expected theoretical intensity ratio as the nearest match, and 69% agreed with the expected ratio within the experimental 1σ uncertainty.

Figure 6a)-c) visualizes the ladder-type transitions observed in measurements #1, #2 and #3 as black sticks, while the transitions predicted by ExoMol [36] are plotted as inverted red sticks indicating their Einstein $A$-coefficients. Some overall common patterns can be observed between the measurement and predictions, but no clear one-to-one correspondence exists that would allow for unambiguous assignments of the observed lines. A reason for predicted lines apparently being absent from the experiment might be interference with the congested Doppler-broadened background. Even using the restrictions on $J$-numbers imposed by combination differences, the experimental intensity ratios, and patterns in transition wavenumber and intensity common between the experiment and predictions, it was not possible to make confident assignments. When possible, we selected the most likely candidates from the predictions and we list these in Table A 1 – A3 in the Appendix for the three pump transitions, together with the experimental line centers, integrated absorption, polarization-

dependent intensity ratios and upper state term values, calculated using our corrected pump transition frequencies (Section 4.2), where the uncertainty is dominated by the uncertainty in the ground state term value. For lines forming combination differences, we report the weighted mean of the term values from each pump-probe combination. For lines measured with the CW probe, we report the center wavenumbers, intensity ratios and term values obtained from the CW measurement.

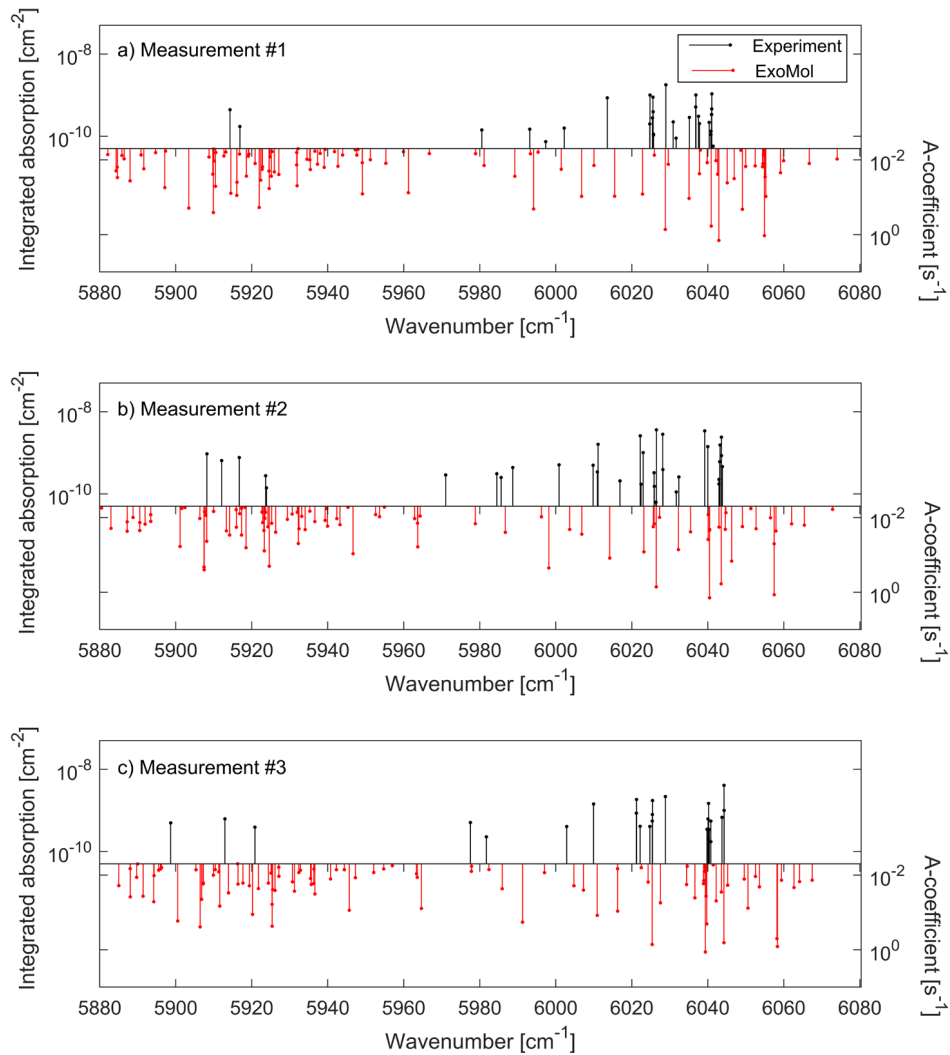

Figure 6. Detected ladder-type transitions (black sticks) and the corresponding ExoMol predictions (red sticks) plotted inverted in a) measurement #1, b) measurement #2 and c) measurement #3. The y-axes show the integrated absorption retrieved from the fits for the measurement, and the Einstein $A$-coefficients for the predictions.

Based on the assignments, the predictions generally appear to be shifted from the experiment towards higher wavenumbers by ∼10 cm$^{-1}$. We stress, however, that the assignments are merely suggestions of the most likely candidates. For three observed lines (forming one combination difference with a term value of 9144.808125(3) cm$^{-1}$) two assignments seemed plausible and thus we include both alternatives. In general, we observed more lines of similar intensity than predicted by ExoMol and 6 predicted transitions are assigned to more than one experimental line. Note, however, that we can in many cases report the upper state $J$-numbers confirmed either by combination differences or by intensity ratios measured with the CW laser, and this is indicated in the last column by "CD" and "CW", respectively. Table A 3 also contains the line measured with the CW probe only, which is flagged accordingly. It forms a combination difference with two lines in Table A 1 and Table A 2 which are flagged as "CD&CW" and the term value is calculated from the CW measurement. Note also that the vibrational labels use the TROVE [36] vibrational quantum numbers, which are not directly comparable to the normal mode labeling conventions used elsewhere in this work and works cited herein.

**4.5. V-type assignment**

We found and fitted 18 V-type transitions in the comb-OODR spectra, most of which were observed in more than one measurement sets, as the lower state was common for all three pump transitions. We assigned 14 of these to the calculated line list of Mraidi *et al.* [33], by comparing to predicted transitions from the state depleted by the pump. The remaining 4 V-type lines had more than one possible assignment and the polarization intensity ratios were not sufficient to resolve the ambiguity, since all plausible candidates belong to the same branch. Figure 7 shows the discrepancies in line center frequencies of the assigned V-type transitions to the theoretical line list of Mraidi *et al.* [33] (red circles) and the assigned experimental line list of Ben Fathallah *et al.* [30] (black squares). The standard deviation of the scatter is 2 GHz and 66 MHz for Refs. [33] and [30], respectively. The agreement with Ben Fathallah *et al.* [30] is within 30 MHz except for the two lines at low frequencies, which are strongly overlapping in the Doppler-broadened regime. This agreement seems reasonable considering that the Ben Fathallah line list was obtained from a Doppler-broadened FTS spectrum with 0.011 cm$^{-1}$ (330 MHz) resolution and estimated a typical uncertainty of 0.001 cm$^{-1}$ (30 MHz) [30].

We also assigned 14 of the V-type transitions to the ExoMol database [36]. The choice of assignments was aided by matching the *J*-numbers to the assignments based on Mraidi *et al.* [33]. However, the wavenumber agreement with ExoMol was much worse than with Ref. [33], and we therefore consider the ExoMol assignments less certain. Since the standard deviation of the discrepancies with ExoMol is 35 GHz, these residuals are not plotted in Figure 7.

Table A 4 lists the retrieved center wavenumbers of the V-type lines compared to the predictions of Ref. [33], whenever assignment was possible. Here the vibrational assignment is given in terms of normal modes, as opposed to the ladder-type assignments. The assignments to ExoMol are presented in Table A 5, where the vibrational states are again given in terms of the TROVE quantum numbers [36]. The reported wavenumbers and uncertainties are obtained from the weighted means of all observations, where the last column indicates in which measurement(s) the line was observed.

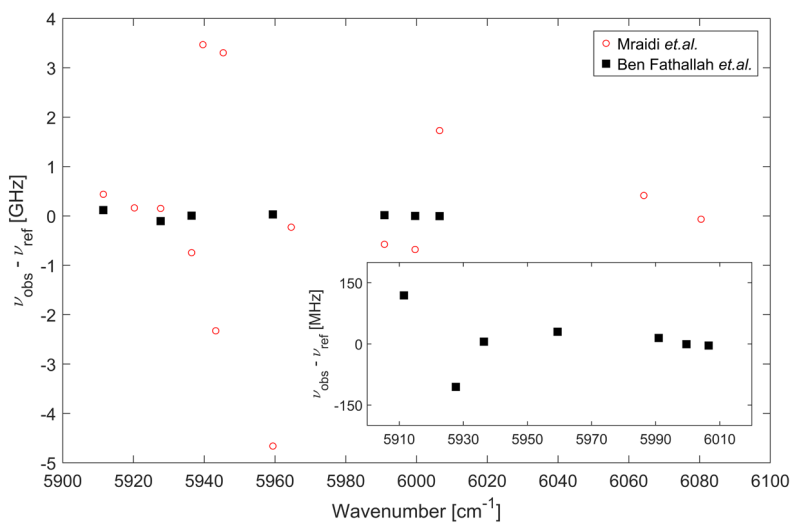

Figure 7. A comparison of retrieved line centers of the assigned V-type transitions to those of the calculated line list of Ref. [33] (red open circles) and the assigned experimental line list of Ref. [30] (black solid squares). The inset shows a zoom of the comparison to Ref. [30].

## 5. Conclusions

We performed the first comb-based OODR measurements of ethylene reaching the unexplored states in the 9000 cm$^{-1}$ range. The comb probe was complemented by a comb-referenced CW probe to improve sensitivity and frequency accuracy for selected OODR lines. We report 90 $C_2H_4$ OODR hot-band transitions between $J = 3 – 5$ states in the $\nu_9$ vibrational mode and final states in the 9030-9180 cm$^{-1}$ range. Assignment to predictions from the ExoMol database [36] proved challenging due to the complexity of the spectrum and large discrepancies with the experiment, but we report tentative assignments for 28 lines. In addition to combination differences, we measured line intensity ratios at parallel and perpendicular relative pump-probe polarization to extract information about the final $J$-numbers of transitions, however the absorption sensitivity in the comb spectra was often insufficient to distinguish the P- and R-branch transitions. The use of a CW probe was highly useful for measuring specific ladder-type lines with improved signal-to-noise ratio, allowing for conclusive branch assignment and improved frequency accuracy. Using combination differences and measurements of intensity ratios with the CW probe, we could confidently assign the upper state $J$-numbers of 34 transitions. Better theoretical predictions are needed for further assignments of the hot-band transitions in this range.

From the splitting of the ladder-type transitions and combination differences, we also determined frequencies of the three pump transitions with 1 order of magnitude improved accuracy compared to line lists from HITRAN2020 and Ben Fatallah *et al*. [21]. Finally, we observed 18 V-type transitions from the ground state in the 5910-6080 cm$^{-1}$ range, and we could assign 14 of them to the calculated line list of Mraidi *et al*. [33] and 14 to ExoMol [36], where the agreement was an order of magnitude better with the former. The frequency agreement with the experimental line list of Ben Fatallah *et al*. [30] was on the 30 MHz level, i.e. within the experimental uncertainties of that work.

# Appendix

Table A 1. Parameters of ladder-type transitions detected when pumping the $^RP_{0,4}(4,A_g)$ transition from the state $(n_1,n_2,n_3,n_4,n_5,n_6,n_7,n_8,n_9,n_{10},n_{11},n_{12},C_v,J,C_{tot},K) = (0,0,0,1,0,0,0,0,0,0,0,0,B_{2u},3,A_u,3)$ using the quantum numbers of ExoMol [36]. Here $n_i$ are the TROVE vibrational quantum numbers, $C_v$ is the vibrational symmetry, $J$ is the total angular momentum, $C_{tot}$ is the ro-vibrational symmetry and $K$ is the projection on the molecular z-axis. The columns show the experimental transition wavenumbers, integrated absorption, polarization-dependent intensity ratios, upper state term values, and the proposed assignments to ExoMol when possible. The last column indicates if the upper state $J$-number has been confirmed using combination differences (CD) or CW measurements (CW) or a combination of both.

| Transition wavenumber [cm$^{-1}$] | Integrated absorption [10$^{-9}$ cm$^{-2}$] | Integrated absorption ratio $\alpha_\parallel/\alpha_\perp$ | Upper state term value [cm$^{-1}$] | Upper state ($J,K,C_{tot}$) | Upper vibrational state | Flag |
|---|---|---|---|---|---|---|
| 5914.286817(8) | 27(2) | 1.1(1) | 9033.53581(7) | (4,4,A$_g$) | 0 0 2 0 1 0 0 0 0 0 0 0 B$_{3g}$ | CD |
| 5916.87890(4) | 10(3) | 1.1(4) | 9036.12802(7) | (4,4,A$_g$) | 0 0 0 1 2 0 0 0 0 0 0 0 A$_g$ | CD&CW |
| 5980.52927(2) | 8(1) | 1.0(2) | 9099.77832(7) | - | - | - |
| 5993.14221(4) | 9(2) | 0.7(2) | 9112.39113(7) | (4,4,A$_g$) | 0 2 0 0 1 0 0 0 0 0 0 0 A$_g$ | CD |
| 5997.34406(4) | 4(1) | 1.4(6) | 9116.59314(7) | (4,4,A$_g$) | 0 2 0 0 1 0 0 0 0 0 0 0 A$_g$ | CD |
| 6002.21175(1) | 9(1) | 0.59(9) | 9121.46080(7) | - | - | - |
| 6013.543722(9) | 52(4) | 1.2(1) | 9132.79277(7) | - | - | - |
| 6024.73124(3) | 12(2) | 1.2(3) | 9143.98029(7) | - | - | - |
| 6024.787277(4) | 60(3) | 1.23(4) | 9144.03633(7) | - | - | - |
| 6025.425864(8) | 17(1) | 1.4(1) | 9144.67492(7) | - | - | - |
| 6025.55908(2) | 53(7) | 1.1(2) | 9144.80812(7) | (4,4,A$_g$) | 0 2 0 0 1 0 0 0 0 0 0 0 A$_g$ | CD |
| | | | | (4,4,A$_g$) | 1 0 1 0 1 0 0 0 1 0 0 0 B$_{3g}$ | CD |
| 6025.604318(8) | 24(2) | 1.1(1) | 9144.85337(7) | - | - | - |
| 6025.68400(1) | 6.7(8) | 1.3(2) | 9144.93305(7) | - | - | - |
| 6025.69955(2) | 6.4(8) | 1.2(2) | 9144.94860(7) | - | - | - |
| 6028.965713(6) | 107(6) | 1.23(7) | 9148.21477(7) | - | - | - |
| 6030.892769(8) | 13.4(9) | 0.76(6) | 9150.14181(7) | (3,2,A$_g$) | 1 0 1 0 1 0 0 0 1 0 0 0 B$_{3g}$ | - |
| 6031.64704(3) | 5.4(9) | 0.6(2) | 9150.89608(7) | - | - | - |
| 6035.14389(2) | 17(2) | 0.8(1) | 9154.39294(7) | (3,2,A$_g$) | 0 0 0 1 2 0 0 0 0 0 0 0 B$_{3g}$ | - |
| 6036.809986(6) | 31(2) | 1.16(8) | 9156.05905(7) | (4,-,A$_g$) | - | CD |
| 6036.851293(6) | 60(3) | 1.10(6) | 9156.10036(7) | (4,4,A$_g$) | 1 0 1 0 1 0 0 0 1 0 0 0 B$_{3g}$ | CD |
| 6037.54163(2) | 18(2) | 1.2(2) | 9156.79068(7) | - | - | - |
| 6037.79627(1) | 12(1) | 1.0(1) | 9157.04531(7) | (4,-,A$_g$) | - | CD |
| 6040.34870(1) | 13(1) | 1.1(2) | 9159.59775(7) | (4,-,A$_g$) | - | CD |
| 6040.80569(1) | 6.6(9) | 1.1(2) | 9160.05474(7) | - | - | - |
| 6040.83375(1) | 7.9(8) | 1.2(1) | 9160.08280(7) | - | - | - |
| 6040.960037(5) | 20(1) | 1.27(7) | 9160.20909(7) | - | - | - |
| 6040.995315(7) | 20(1) | 1.01(6) | 9160.24438(7) | (4,-,A$_g$) | - | CD |
| 6041.022271(4) | 28(1) | 1.08(4) | 9160.27134(7) | (4,-,A$_g$) | - | CD |
| 6041.056365(2) | 64(3) | 1.06(2) | 9160.30543(7) | (4,4,A$_g$) | 0 0 0 1 2 0 0 0 0 0 0 0 B$_{3g}$ | CD |
| 6041.39073(1) | 3.4(5) | 1.0(2) | 9160.63979(7) | - | - | - |

Table A 2. Parameters of ladder-type transitions detected when pumping the $^RQ_{0,4}(4,A_g)$ transition from the state $(n_1,n_2,n_3,n_4,n_5,n_6,n_7,n_8,n_9,n_{10},n_{11},n_{12},C_v,J,C_{tot},K) = (0,0,0,1,0,0,0,0,0,0,0,0,B_{2u},4,A_u,1)$ using the quantum numbers of ExoMol [36]. Here $n_i$ are the TROVE vibrational quantum numbers, $C_v$ is the vibrational symmetry, $J$ is the total angular momentum, $C_{tot}$ is the ro-vibrational symmetry and $K$ is the projection on the molecular z-axis. The columns show the experimental transition wavenumbers, integrated absorption, polarization-dependent intensity ratios, upper state term values, and the proposed assignments to ExoMol when possible. The last column indicates if the upper state $J$-number has been confirmed using combination differences (CD) or CW measurements (CW) or a combination of both.

| Transition wavenumber [cm$^{-1}$] | Integrated absorption [10$^{-9}$ cm$^{-2}$] | Integrated absorption ratio $\alpha_\parallel/\alpha_\perp$ | Upper state term value [cm$^{-1}$] | Upper state $(J,K,C_{tot})$ | Upper vibrational state | Flag |
|---|---|---|---|---|---|---|
| 5908.20741(4) | 57(27) | 2(1) | 9036.12802(7) | (4,4,A$_g$) | 0 0 0 1 2 0 0 0 0 0 0 0 A$_g$ | CD&CW |
| 5912.07682(4) | 39(7) | 1.4(3) | 9039.99751(8) | - | - | - |
| 5916.71680(3) | 46(8) | 1.2(3) | 9044.63749(7) | - | - | - |
| 5923.65118(3) | 17(3) | 0.6(2) | 9051.57187(7) | - | - | - |
| 5923.86661(3) | 8(3) | 1.7(9) | 9051.78730(7) | - | - | - |
| 5971.03168(4) | 17(4) | 0.8(3) | 9098.95237(8) | - | - | - |
| 5984.47038(2) | 19(4) | 3(1) | 9112.39113(7) | (4,4,A$_g$) | 0 2 0 0 1 0 0 0 0 0 0 0 A$_g$ | CD |
| 5985.61604(4) | 15(4) | 2.3(9) | 9113.53673(8) | - | - | - |
| 5988.67249(8) | 26(14) | 2(1) | 9116.59314(7) | (4,4,A$_g$) | 0 2 0 0 1 0 0 0 0 0 0 0 A$_g$ | CD |
| 6000.85341(1) | 31(2) | 2.0(2) | 9128.77410(7) | - | - | - |
| 6009.81383(2) | 30(4) | 1.0(2) | 9137.73454(7) | (5,4,A$_g$) | 0 2 0 0 1 0 0 0 0 0 0 0 A$_g$ | - |
| 6010.92863(2) | 21(2) | 0.65(8) | 9138.84932(7) | - | - | - |
| 6011.119498(7) | 96(5) | 0.61(3) | 9139.04019(7) | - | - | - |
| 6016.88742(6) | 12(4) | 3(1) | 9144.80812(7) | (4,4,A$_g$) | 0 2 0 0 1 0 0 0 0 0 0 0 A$_g$ | CD |
| | | | | (4,4,A$_g$) | 1 0 1 0 1 0 0 0 1 0 0 0 B$_{3g}$ | CD |
| 6022.221117(4) | 155(6) | 0.54(1) | 9150.14181(7) | (3,2,A$_g$) | 1 0 1 0 1 0 0 0 1 0 0 0 B$_{3g}$ | - |
| 6022.42012(3) | 10(2) | 0.7(2) | 9150.34081(8) | - | - | - |
| 6022.975388(6) | 61(3) | 0.61(3) | 9150.89608(7) | - | - | - |
| 6025.85978(1) | 20(2) | 0.48(6) | 9153.78048(7) | - | - | - |
| 6025.92331(3) | 9(2) | 0.5(1) | 9153.84401(7) | - | - | - |
| 6026.33840(2) | 3.8(8) | 0.6(2) | 9154.25909(7) | - | - | - |
| 6026.472250(3) | 217(8) | 0.556(8) | 9154.39294(7) | (3,2,A$_g$) | 0 0 0 1 2 0 0 0 0 0 0 0 B$_{3g}$ | - |
| 6028.135001(4) | 169(7) | 0.75(1) | 9156.05569(7) | - | - | - |
| 6028.17972(8) | 23(10) | 2(1) | 9156.10036(7) | (4,4,A$_g$) | 1 0 1 0 1 0 0 0 1 0 0 0 B$_{3g}$ | CD |
| 6031.69594(2) | 7(2) | 0.5(2) | 9159.61663(7) | - | - | - |
| 6032.38476(4) | 16(2) | 1.7(4) | 9160.30543(7) | (4,4,A$_g$) | 0 0 0 1 2 0 0 0 0 0 0 0 B$_{3g}$ | CD |
| 6039.171863(6) | 204(9) | 0.75(3) | 9167.09256(7) | - | - | - |
| 6039.98773(2) | 85(9) | 0.7(1) | 9167.90842(7) | - | - | - |
| 6042.93114(3) | 10(2) | 0.9(3) | 9170.85184(8) | - | - | - |
| 6042.93746(3) | 13(2) | 1.0(2) | 9170.85815(7) | - | - | - |
| 6043.11662(2) | 36(4) | 0.7(1) | 9171.03731(7) | - | - | - |
| 6043.203949(6) | 93(4) | 0.74(3) | 9171.12464(7) | - | - | - |
| 6043.574095(4) | 144(6) | 0.77(2) | 9171.49479(7) | - | - | - |
| 6043.584678(9) | 51(3) | 0.71(4) | 9171.50537(7) | - | - | - |
| 6043.82159(3) | 28(3) | 0.7(1) | 9171.74228(8) | - | - | - |

Table A 3. Parameters of ladder-type transitions detected when pumping the $^RR_{0,4}(4,A_g)$ transition from the state $(n_1,n_2,n_3,n_4,n_5,n_6,n_7,n_8,n_9,n_{10},n_{11},n_{12},C_v,J,C_{tot},K) = (0,0,0,1,0,0,0,0,0,0,0,0,B_{2u},5,A_u,5)$ using the quantum numbers of ExoMol [36]. Here $n_i$ are the TROVE vibrational quantum numbers, $C_v$ is the vibrational symmetry, $J$ is the total angular momentum, $C_{tot}$ is the ro-vibrational symmetry and $K$ is the projection on the molecular z-axis. The columns show the experimental transition wavenumbers, integrated absorption, polarization-dependent intensity ratios, upper state term values, and the proposed assignments to ExoMol when possible. The last column indicates if the upper state $J$-number has been confirmed using combination differences (CD) or CW measurements (CW) or a combination of both.

| Transition wavenumber [cm$^{-1}$] | Integrated absorption [10$^{-9}$ cm$^{-2}$] | Integrated absorption ratio $\alpha_\parallel/\alpha_\perp$ | Upper state term value [cm$^{-1}$] | Upper state $(J,K,C_{tot})$ | Upper vibrational state | Flag |
|---|---|---|---|---|---|---|
| 5898.67093(3) | 30(7) | 1.3(4) | 9033.53581(7) | (4,4,A$_g$) | 0 0 2 0 1 0 0 0 0 0 0 0 B$_{3g}$ | CD |
| 5901.2634675(1) | - | 1.52(5) | 9036.12802(7) | (4,4,A$_g$) | 0 0 0 1 2 0 0 0 0 0 0 0 A$_g$ | CW only |
| 5912.9344311(1) | 37(5) | 1.29(4) | 9047.79898(7) | (6,6,A$_g$) | 0 0 2 0 1 0 0 0 0 0 0 0 B$_{3g}$ | CW |
| 5920.8322341(1) | 23(2) | 1.28(4) | 9055.69679(7) | (6,6,A$_g$) | 0 0 0 1 2 0 0 0 0 0 0 0 A$_g$ | CW |
| 5977.5268(1) | 30(16) | 2(1) | 9112.39113(7) | (4,4,A$_g$) | 0 2 0 0 1 0 0 0 0 0 0 0 A$_g$ | CD |
| 5981.72859(2) | 14(2) | 1.4(2) | 9116.59314(7) | (4,4,A$_g$) | 0 2 0 0 1 0 0 0 0 0 0 0 A$_g$ | CD |
| 6002.869990(6) | 24(1) | 0.50(3) | 9137.73454(7) | (5,4,A$_g$) | 0 2 0 0 1 0 0 0 0 0 0 0 A$_g$ | - |
| 6009.943572(3) | 85(4) | 1.43(3) | 9144.80812(7) | (4,4,A$_g$) | 0 2 0 0 1 0 0 0 0 0 0 0 A$_g$ | CD |
|  |  |  |  | (4,4,A$_g$) | 1 0 1 0 1 0 0 0 1 0 0 0 B$_{3g}$ | CD |
| 6021.19453(2) | 51(7) | 1.7(3) | 9156.05905(7) | (4,-,A$_g$) | - | CD |
| 6021.235804(4) | 110(5) | 1.41(4) | 9156.10036(7) | (4,4,A$_g$) | 1 0 1 0 1 0 0 0 1 0 0 0 B$_{3g}$ | CD |
| 6022.180758(9) | 24(1) | 1.32(7) | 9157.04531(7) | (4,-,A$_g$) | - | CD |
| 6024.733200(8) | 24(1) | 1.35(8) | 9159.59775(7) | (4,-,A$_g$) | - | CD |
| 6025.37986(2) | 33(5) | 1.4(3) | 9160.24438(7) | (4,-,A$_g$) | - | CD |
| 6025.406797(6) | 47(3) | 1.44(7) | 9160.27134(7) | (4,-,A$_g$) | - | CD |
| 6025.440881(3) | 104(5) | 1.47(2) | 9160.30543(7) | (4,4,A$_g$) | 0 0 0 1 2 0 0 0 0 0 0 0 B$_{3g}$ | CD |
| 6028.833430(3) | 129(6) | 1.28(3) | 9163.69798(7) | - | - | - |
| 6039.7881914(1) | 21(2) | 1.29(9) | 9174.65274(7) | (6,-,A$_g$) | - | CW |
| 6040.041809(7) | 37(2) | 1.24(6) | 9174.90636(7) | - | - | - |
| 6040.12485(1) | 12(1) | 1.3(1) | 9174.98940(7) | - | - | - |
| 6040.192083(4) | 88(4) | 1.31(3) | 9175.05664(7) | - | - | - |
| 6040.39800(1) | 20(2) | 1.7(2) | 9175.26255(7) | - | - | - |
| 6040.78911(2) | 10(2) | 1.1(3) | 9175.65366(7) | - | - | - |
| 6040.794621(7) | 33(2) | 1.19(9) | 9175.65917(7) | - | - | - |
| 6043.724355(5) | 41(2) | 1.25(4) | 9178.58891(7) | - | - | - |
| 6044.25967246(1) | 243(12) | 1.26(3) | 9179.12423(7) | (6,-,A$_g$) | - | CW |
| 6044.28714(1) | 60(6) | 1.3(1) | 9179.15170(7) | - | - | - |

Table A 4. The experimental V-type line center wavenumbers together with the transition wavenumbers, intensities and assignments from the calculated line list of Mraidi *et al.* [33]. Here, $J$ is the total angular momentum, $C_{tot}$ the ro-vibrational symmetry, and $N$ the ranking number. The vibrational label is on the form $(n_1,n_2,n_3,n_4,n_5,n_6,n_7,n_8,n_9,n_{10},n_{11},n_{12},C_v)$ where $n_i$ are normal mode vibrational quantum numbers and $C_v$ is the vibrational symmetry. The last column states in which measurements the V-type transition was observed. Reported line centers are the weighted means of all observations.

| Experimental transition wavenumber [cm$^{-1}$] | Calculated transition wavenumber [cm$^{-1}$] | Calculated line intensity [cm/mol] | Upper state $(J,C_{tot},N)$ | Upper state vibrational state | Observed in measurement(s) |
|---|---|---|---|---|---|
| 5911.473849(7) | 5911.45923 | 3.971·10$^{-23}$ | (3,A$_u$,219) | 0 1 1 0 0 0 0 0 0 0 1 0 B$_{1u}$ | #1,#2,#3 |
| 5920.254425(9) | 5920.249 | 2.853·10$^{-23}$ | (3,A$_u$,220) | 0 1 0 0 0 1 0 0 1 0 0 0 B$_{1u}$ | #1,#2,#3 |
| 5927.669751(8) | 5927.66469 | 5.051·10$^{-23}$ | (5,A$_u$,339) | 0 1 1 0 0 0 0 0 0 0 1 0 B$_{1u}$ | #1,#2,#3 |
| 5929.26754(2) | - | - | - | - | #1 |
| 5932.78686(1) | - | - | - | - | #2,#3 |
| 5936.422652(5) | 5936.44748 | 3.293·10$^{-23}$ | (5,A$_u$,340) | 0 1 0 0 0 1 0 0 1 0 0 0 B$_{1u}$ | #1,#2,#3 |
| 5939.61745(1) | 5939.50175 | 1.065·10$^{-23}$ | (3,A$_u$,399) | 0 0 0 0 0 0 0 0 0 2 0 3 B$_{1u}$ | #1,#3 |
| 5943.249792(6) | 5943.32741 | 3.006·10$^{-23}$ | (3,A$_u$,57) | 1 0 0 0 0 0 0 0 0 0 1 0 B$_{1u}$ | #1,#2,#3 |
| 5945.36819(2) | 5945.25799 | 4.893·10$^{-24}$ | (3,A$_u$,111) | 0 0 0 4 0 0 1 1 0 0 0 0 B$_{1u}$ | #1,#3 |
| 5955.77755(1) | - | - | - | - | #3 |
| 5959.402999(5) | 5959.55846 | 3.856·10$^{-23}$ | (5,A$_u$,90) | 1 0 0 0 0 0 0 0 0 0 1 0 B$_{1u}$ | #1,#2,#3 |
| 5964.580252(9) | 5964.58787 | 3.529·10$^{-23}$ | (4,A$_u$,281) | 0 0 0 0 0 0 0 0 1 0 0 2 B$_{3u}$ | #2,#3 |
| 5971.385584(9) | - | - | - | - | #1,#2,#3 |
| 5990.930702(5) | 5990.94989 | 6.072·10$^{-23}$ | (3,A$_u$,59) | 0 0 0 0 1 0 0 0 0 0 1 0 B$_{3u}$ | #1,#2,#3 |
| 5999.600548(7) | 5999.62323 | 1.673·10$^{-22}$ | (4,A$_u$,72) | 0 0 0 0 1 0 0 0 0 0 1 0 B$_{3u}$ | #1,#2,#3 |
| 6006.522086(6) | 6006.46443 | 1.301·10$^{-22}$ | (5,A$_u$,92) | 0 0 0 0 1 0 0 0 0 0 1 0 B$_{3u}$ | #1,#2,#3 |
| 6064.260529(9) | 6064.24673 | 3.331·10$^{-23}$ | (3,A$_u$,226) | 1 1 0 0 0 0 0 0 0 0 0 1 B$_{1u}$ | #1,#2 |
| 6080.42774(1) | 6080.42996 | 4.112·10$^{-23}$ | (5,A$_u$,351) | 1 1 0 0 0 0 0 0 0 0 0 1 B$_{1u}$ | #1,#2 |

Table A 5. The experimental V-type line center wavenumbers together with the transition wavenumbers, intensities and assignments from ExoMol [36]. Here, $J$ is the total angular momentum, $K$ is the projection on the molecular z-axis and $C_{tot}$ is the ro-vibrational symmetry. The vibrational label is on the form $(n_1,n_2,n_3,n_4,n_5,n_6,n_7,n_8,n_9,n_{10},n_{11},n_{12},C_v)$ where $n_i$ are the TROVE vibrational quantum numbers and $C_v$ is the vibrational symmetry. The last column states in which measurements the V-type transition was observed. Reported line centers are the weighted means of all observations.

| Experimental transition wavenumber [cm$^{-1}$] | Calculated transition wavenumber [cm$^{-1}$] | Calculated line intensity [cm/mol] | Upper state $(J,K,C_{tot})$ | Upper state vibrational state | Observed in measurement(s) |
|---|---|---|---|---|---|
| 5911.473849(7) | 5911.548584 | 3.235·10$^{-23}$ | (3,3,A$_u$) | 0 0 1 0 0 0 1 1 0 0 0 0 B$_{1u}$ | #1,#2,#3 |
| 5920.254425(9) | 5921.772501 | 4.745·10$^{-23}$ | (3,3,A$_u$) | 1 0 0 1 0 0 0 1 0 0 0 0 B$_{1u}$ | #1,#2,#3 |
| 5927.669751(8) | 5927.68416 | 3.884·10$^{-23}$ | (5,5,A$_u$) | 0 0 1 0 0 0 1 1 0 0 0 0 B$_{1u}$ | #1,#2,#3 |
| 5929.26754(2) | 5929.450289 | 5.462·10$^{-24}$ | (3,3,A$_u$) | 1 0 0 0 0 1 2 1 0 0 0 0 B$_{2u}$ | #1 |
| 5932.78686(1) | - | - | - | - | #2,#3 |
| 5936.422652(5) | 5937.877975 | 6.13·10$^{-23}$ | (5,5,A$_u$) | 1 0 0 1 0 0 0 1 0 0 0 0 B$_{1u}$ | #1,#2,#3 |
| 5939.61745(1) | 5937.49087 | 5.562·10$^{-24}$ | (3,3,A$_u$) | 1 0 0 0 0 1 0 0 1 2 0 0 B$_{1u}$ | #1,#3 |
| 5943.249792(6) | 5944.664535 | 4.384·10$^{-23}$ | (3,3,A$_u$) | 0 2 0 0 0 0 0 0 0 0 0 0 B$_{1u}$ | #1,#2,#3 |
| 5945.36819(2) | - | - | - | - | #1,#3 |
| 5955.77755(1) | 5955.934982 | 1.371·10$^{-23}$ | (3,3,A$_u$) | 0 0 0 1 0 1 0 1 0 0 0 0 B$_{2u}$ | #3 |
| 5959.402999(5) | 5960.762151 | 5.673·10$^{-23}$ | (5,5,A$_u$) | 0 2 0 0 0 0 0 0 0 0 0 0 B$_{1u}$ | #1,#2,#3 |
| 5964.580252(9) | 5964.706227 | 3.052·10$^{-23}$ | (4,1,A$_u$) | 0 0 0 1 0 1 0 1 0 0 0 0 B$_{2u}$ | #2,#3 |
| 5971.385584(9) | - | - | - | - | #1,#2,#3 |
| 5990.930702(5) | 5992.721804 | 7.849·10$^{-23}$ | (3,3,A$_u$) | 0 0 0 2 0 0 0 0 0 0 0 0 B$_{2u}$ | #1,#2,#3 |
| 5999.600548(7) | - | - | - | - | #1,#2,#3 |
| 6006.522086(6) | 6008.245482 | 1.312·10$^{-22}$ | (5,5,A$_u$) | 0 0 0 2 0 0 0 0 0 0 0 0 B$_{2u}$ | #1,#2,#3 |
| 6064.260529(9) | 6066.328667 | 2.674·10$^{-23}$ | (3,3,A$_u$) | 1 0 0 1 0 0 0 0 1 0 0 0 B$_{1u}$ | #1,#2 |
| 6080.42774(1) | 6082.406459 | 3.376·10$^{-23}$ | (5,5,A$_u$) | 1 0 0 1 0 0 0 0 1 0 0 0 B$_{1u}$ | #1,#2 |

## Acknowledgements

The authors thank Jonathan Tennyson and Sergey Yurchenko for providing the ground state term value from their ongoing MARVEL project.

This project is supported by the Knut and Alice Wallenberg Foundation (grant: KAW 2020.0303), and the Swedish Research Council (grant: 2020-00238). Y. C. acknowledges the Wenner Gren Foundation (grant: UPD2024-0201). K.K.L. acknowledges funding from the U.S. National Science Foundation (grant: CHE-2108458) and the Wenner Gren Foundation (grant: GFOv2024-0010).